# Modeling Transient Electromagnetic Logging Using Sine Transform and Levin-Sidi Extrapolation

Feng-Feng Li    cj21701@163.com

*Abstract*—Transient electromagnetic logging (TEL) is of great significance for detecting deep formation structures outside the wellbore. Forward modeling using analytical solutions requires frequency-domain calculations followed by time-domain conversion. For electromagnetic field computations in horizontally stratified media, mature Hankel transform methods already enable precise solutions. However, for cylindrically stratified media, oscillatory integrals involving cosine functions must be evaluated. This study employs the sine transform to achieve frequency-to-time domain conversion for TEL, proposing a new set of 125-point sine transform filter coefficients with enhanced computational accuracy. For frequency-domain calculations in cylindrically stratified media, the Gauss-Legendre quadrature rule combined with Levin-Sidi extrapolation is adopted to accelerate the convergence of partial sum sequences toward the true integral value. The accuracy of the proposed computational method is verified through numerical experiments in both horizontally layered and cylindrically stratified media.

## 1. Introduction

Transient electromagnetic signals are generated by step-current excitation and detect subsurface features by utilizing the time-decay characteristics of induced eddy current fields in underground media after power cutoff [1], [2]. This method aids in identifying potential subsurface conductors, evaluating formation physical property parameters, and understanding geological structures [3]. In borehole scenarios, the wide spectral characteristics of transient electromagnetic signals endow them with excellent depth penetration capabilities in complex media. This feature enables Transient electromagnetic logging (TEL) to be applied not only for detecting deep formation structures in open-hole wells but also for effectively penetrating high-conductivity casings in production wells to accurately capture reservoir information outside the casings [4], [5].

Directly computing transient electromagnetic wave responses in multi-layered media via time-domain methods not only demands substantial computational resources but also incurs high costs, especially for high-precision applications [6]. Converting frequency-domain data to time-domain data for analyzing time-dependent responses reduces computational load, yet improper implementation of this process may compromise accuracy, while traditional inverse Fourier transform methods remain computationally inefficient and inflexible, limiting both speed and precision [7], [8]. The Laplace inversion method finds extensive applications in computing transient responses. Among them, the Gaver-Stehfest method is easy to implement in engineering scenarios. However, it tends to yield deviations in "long-time" calculations and is thus suitable for "short-time" transient signal computations [9]. Algorithms like Euler and Talbot are also Laplace inversion methods, but their accuracy for "long-time" transient signals still falls short of requirements [10].

Digital filtering techniques have become quite mature in the computational field of Hankel transform [11]. Sine and cosine functions can be transformed into Bessel function forms, so the digital filters for Fourier sine and cosine transforms can be converted into those for Hankel transforms [12]. Appropriate filters are key to enabling the effective application of sine and cosine transforms in transient electromagnetic field calculations, with fewer filtering coefficients resulting in faster calculation speeds.

For frequency-domain electromagnetic field calculations in horizontally stratified media, precise solutions can be achieved using existing Hankel transforms [13]. However, for cylindrically stratified media, high-precision integration of oscillatory functions containing cosine terms is required [14]. To address time-domain conversion and frequency-domain computation challenges, this study employs the Gauss-Legendre quadrature method combined with Levin-Sidi extrapolation to solve for frequency-domain electromagnetic logging responses. Additionally, a set of 125-point sine filter coefficients is utilized to enable efficient frequency-to-time domain electromagnetic field conversion.

## 2. Integration methods

### 1.1. Time-Domain Integration Based on Sine Transform

In the modeling of the transient electromagnetic method, the construction of time-domain electromagnetic responses is achieved through inverse Fourier transform. The Fourier sine and cosine transforms are defined as [12], [13], [15]:

$$f(v) = \int_0^\infty k(u)\cos(uv)du$$
$$f(v) = \int_0^\infty k(u)\sin(uv)du \tag{1}$$

where $k(u)$ is the kernel function.

The cosine and sine functions can be expressed using Bessel functions of the first kind with order $\pm 1/2$:

$$\begin{matrix}\cos\\ \sin\end{matrix}(uv) = \sqrt{\frac{\pi}{2}}\sqrt{uv}J_{\mp\frac{1}{2}}(uv) \tag{2}$$

Substituting Eq. (2) into Eq. (1) yields:

$$f(v) = \sqrt{\frac{\pi}{2}}\int_0^\infty k(u)\sqrt{uv}J_{-\frac{1}{2}}(uv)du$$
$$f(v) = \sqrt{\frac{\pi}{2}}\int_0^\infty k(u)\sqrt{uv}J_{+\frac{1}{2}}(uv)du \tag{3}$$

For sine transform, let $uv = \omega$, $v = t$, leading to:

$$f(t) = \sqrt{\frac{\pi}{2}}\int_0^\infty k(\frac{\omega}{t})\sqrt{\omega}J_{+\frac{1}{2}}(\omega)d\omega \tag{4}$$

Exponential sampling is performed on $\omega$ and $t$, let $\omega = e^{m\cdot s}$, $t = e^{n\cdot s}$, and substituting these into the above equation gives:

$$e^{n\cdot s}f(e^{n\cdot s}) = \sqrt{\frac{\pi}{2}}\int_{-\infty}^\infty k(e^{-s(n-m)})\sqrt{e^{m\cdot s}}se^{m\cdot s}J_{+\frac{1}{2}}(e^{m\cdot s})dm \tag{5}$$

This equation can be expressed in the form of a convolution.:

$$Y(n) = \int_{-\infty}^\infty X(n-m)H(m)dm \tag{6}$$

where

$$X(n-m) = k(e^{-s(n-m)}) \tag{7}$$

$$Y(n) = \sqrt{\frac{2}{\pi}}e^{sn}f(e^{sn}) \tag{8}$$

$$H(m) = \sqrt{e^{sm}}(se^{sm})J_{+\frac{1}{2}}(e^{sm}) \tag{9}$$

The above equation is discretely approximated as $Y(n) = \sum_{m=-\infty}^{\infty} X(n-m)H(m)$. By constructing and solving a system of linear equations based on the known samples of the input function *X* and output function *Y*, the digital filter coefficients *H* can be obtained. Conversely, given the input function *X* and the digital filter coefficients *H*, the time-domain result *Y* can also be derived. This study employs optimization methods [12] to derive a set of 125-point sine filter coefficients (Appendix), which are subsequently applied for time-domain computations.

The time-domain response can be approximated using digital filters with a finite number of points:

$$\frac{\partial_{H_z(t)}}{\partial_t} = -\sqrt{\frac{2}{\pi t^2}} \sum_{j=-N}^{N} \text{Im}(H_z(\frac{\omega_j}{t})) \cdot h_{j,\sin} \qquad (10)$$

where, $\omega_j = e^{-s \cdot j}$, where *s* is the sampling interval, and $h_{j,\sin}$ is the filter coefficients *H*.

The induced electromotive force on the receiving coil is:

$$V(t) = N_T N_R S \mu_0 I_T \frac{\partial_{H_z(t)}}{\partial_t} \qquad (11)$$

where, *S* is the area of the receiving coil; $N_T$ and $N_R$ are the number of turns in the coils; *I* is the transmitting current intensity.

*1.2. Frequency-Domain Integration Based on Levin-sidi method*

The spatial-domain magnetic field $H_Z$ in cylindrically stratified media is obtained through the inverse Fourier cosine transform of the spectral-domain magnetic field $\tilde{H}_z$ along the wavenumber $k_z$ direction.

$$H_z(\rho, z) = \int_0^\infty \cos(k_z z) \tilde{H}_z(\rho, k_z) dk_z \qquad (12)$$

The integrand of this expression consists of an integral kernel and a cosine function, forming an oscillatory integral with an integration interval of [0, ∞], which can be expressed as:

$$F(r) = \int_0^\infty f(k) g(kr) dk \qquad (13)$$

Where, *f(k)* is the kernel function; *g(kr)* is the cosine function.

For high-accuracy integration of cosine-containing oscillatory functions, we apply Gauss-Legendre quadrature on finite subintervals with Levin-Sidi extrapolation to accelerate partial sum convergence[16].

1) Gauss-Legendre quadrature

The domain of integration is discretized into finite subintervals $[a_i, a_{i+1}]$, where *i* = 0, 1, 2, ..., with $\lim_{i \to 0} a_i = \infty$, and $a_{i+1} - a_i = \pi$. The integral over each subinterval is evaluated using Gauss-Legendre quadrature:

$$u_i = \int_{a_i}^{a_{i+1}} f(k) g(kr) dk \qquad (14)$$

To enable fast filter-based evaluation, we apply a change of variables to rewrite the preceding equation as:

$$u_i = \int_{a_i}^{a_{i+1}} f(\frac{k}{r}) \frac{g(k)}{r} dk \qquad (15)$$

The Gauss-Legendre quadrature is conventionally applied to [-1, 1]. Applying substitution $t = \frac{2k - (a+b)}{b-a}$, yields $k = \frac{b-a}{2} t + \frac{a+b}{2}$ and $dk = \frac{b-a}{2} dt$, converting the integral to:

$$\int_{-1}^{1} \frac{f\left(\frac{(b-a)t+(a+b)}{2r}\right)}{g\left(\frac{(b-a)t+(a+b)}{2}\right)} \Bigg/ r \cdot \left(\frac{b-a}{2}\right) dt \qquad (16)$$

The integration order $n$ determined according to the desired accuracy level, with the associated quadrature nodes $t_i$ and weights $w_i$ ($i = 0, 1, 2,..., n$) being retrieved from standard Gauss-Legendre tables.

First, evaluate the integrand at the given quadrature nodes $t_i$:

$$h(t_i) = \frac{f\left(\frac{(b-a)t_i+(a+b)}{2r}\right)}{g\left(\frac{(b-a)t_i+(a+b)}{2}\right)} \Bigg/ r \cdot \left(\frac{b-a}{2}\right) \qquad (17)$$

The approximate value $u_i \approx \sum_{i=1}^{n} w_i h(t_i)$ of the subinterval integral is then computed via the Gauss-Legendre quadrature rule.

2) Levin transform

Define the integral sequence $S_n = \sum_{i=0}^{n} u_i$, then formulate the Levin transformation via remainder expansion. The k-th order transformed partial sum $S_n^{(k)}$ is obtained by applying Cramer's rule to derive the Levin transformation formula:

$$S_n^{(k)} = \sum_{j=0}^{k} \pi_n^{(k,j)} \frac{S_{n+j}}{\omega_{n+j}} \Bigg/ \sum_{j=0}^{k} \pi_n^{(k,j)} \frac{1}{\omega_{n+j}} \qquad (18)$$

Where, $\pi_n^{(k,j)} = \prod_{\substack{m=0 \\ m \neq j}}^{k} \frac{1}{x_{n+m}^{-1} - x_{n+j}^{-1}}$; $x_n$ is the extrapolation sequence.

3) W-algorithm of sidi

For efficient Levin transformation computation, we apply Sidi's W-algorithm. Compute numerator/denominator using recursive formula $R_n^{(k+1)} = \frac{R_{n+1}^{(k)} - R_n^{(k)}}{x_{n+k+1}^{-1} - x_n^{-1}}$ ($n \geq 0, k \geq 0$) (initialized with $A_n^{(0)} = \frac{S_n}{\omega_n}$ and $B_n^{(0)} = \frac{1}{\omega_n}$). The accelerated sequence $S_n^{(k)} = \frac{A_n^{(k)}}{B_n^{(k)}}$ is obtained by progressively increasing the order $k$ until convergence to the desired precision is achieved. The final integral result is given by $S_n$. We employs the u-transformation to compute $\omega_n$, $\omega_n = x_n \cdot \Delta S_{n-1} = x_n(S_n - S_{n-1})$, where the sequence is initialized with $\omega_1 = x_1 \cdot S_1$.

Fig.1 shows how Gauss points and extrapolation points affect integration: (a) Fixed 20 extrapolation points; (b) Fixed 20 Gauss points. Results stabilize when both parameters exceed 6. For time-domain accuracy, we use 16 Gauss points and max 10 extrapolation points.

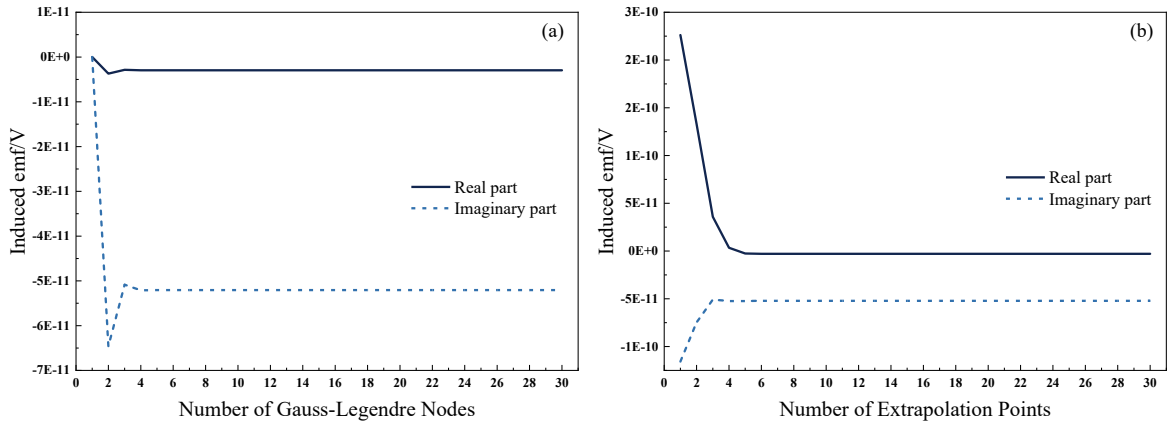

**Fig.1.** Effect of gauss points and extrapolation points on oscillatory integral results

## 3. Numerical examples

### 3.1. Horizontally multilayered medium

For frequency-domain computations, the implementation incorporates a 121-point zeroth-order Hankel transform paired with a 140-point first-order Hankel transform, whereas the time-domain calculations utilize the proposed 125-point sine transform algorithm.

As illustrated in Fig.2, which analyzes relative errors between numerical integration and analytical solutions in homogeneous formations, panel (a) establishes that the relative error consistently remains below 0.1% for resistivity variations spanning 0.1–100 Ω·m at a fixed tool spacing of 1 m, while panel (b) confirms the maintenance of 0.1% accuracy across tool spacing variations up to 10 m under a constant resistivity of 1 Ω·m.

The comparative study of TEL responses in two-layer formations (Fig.3), configured with a transmitter positioned 10 m from the formation boundary and a source spacing of 1 m, demonstrates that the integral solution exhibits superior accuracy in early-time responses ($t < 10^{-4}$ s) while maintaining rigorous agreement with finite-element solutions.

Further analysis of a 60° deviated well penetrating multilayered formations (Fig.4), characterized by resistivity contrasts of 10 Ω·m (high) versus 0.2 Ω·m (low) and layer thicknesses varying from 0.15 to 3 m, reveals that the 125-point sine transform delivers exceptional stability in both transient response calculations and apparent resistivity inversion.

Collectively, the results presented in Fig2-4 validate the 125-point sine transform's capability to achieve sub-0.1% numerical accuracy, robust stability in high-contrast formations, reliable performance in deviated well geometries, and computational efficiency surpassing conventional methods, thereby substantiating its efficacy for transient electromagnetic logging applications.

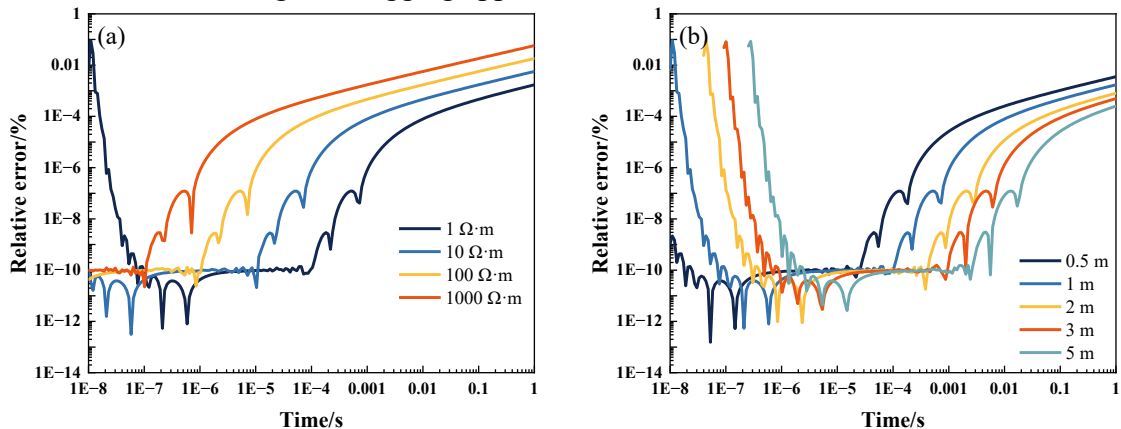

**Fig.2.** Relative errors in homogeneous formation

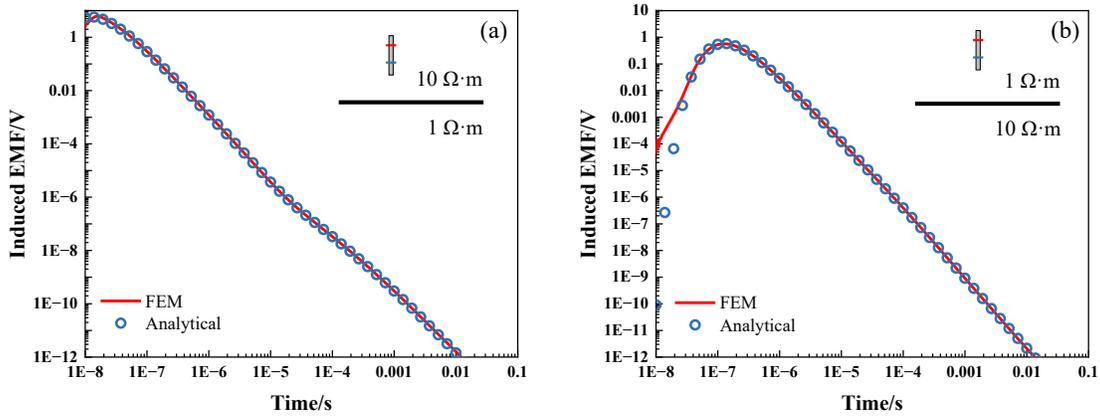

**Fig.3.** Forward modeling in two-layered formations

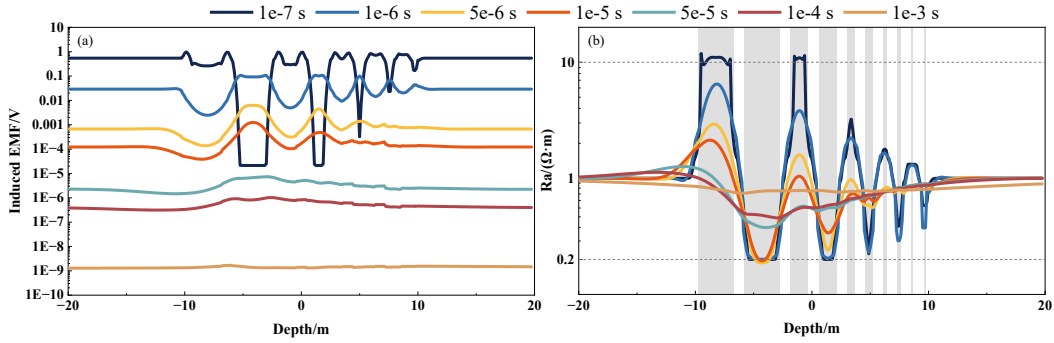

**Fig.4.** Forward modeling in multilayered formations

*3.2. Cylindrically multilayered medium*

Fig.5 illustrates the comparative analysis of TEL responses in cylindrically multilayered medium. The geometric model (left panel) comprises four concentric zones: Medium 1 (0–0.0825 m): Tool mandrel; Medium 2 (0.0825–0.1016 m): Borehole fluid; Medium 3 (0.1016–0.1116 m): Steel casing; Medium 4 (>0.1116 m): Formation. The transmitter and receiver coils have a radius of 0.0965 m.

We investigated two resistivity configurations:
1) Open-hole simulation (Fig. 5a): Conductive mandrel ($1×10^{-7}$ Ω·m), Resistive borehole (1000 Ω·m), Formation/casing at 1 Ω·m.
2) Cased-hole simulation (Fig. 5b): Insulated mandrel (1000 Ω·m), Perfectly conductive casing ($1×10^{-7}$ Ω·m).

The numerical solutions exhibit remarkable consistency between integral (Levin-Sidi frequency-domain + 125-point time-domain) and finite-element methods, confirming the robustness of the proposed algorithms under extreme resistivity contrasts.

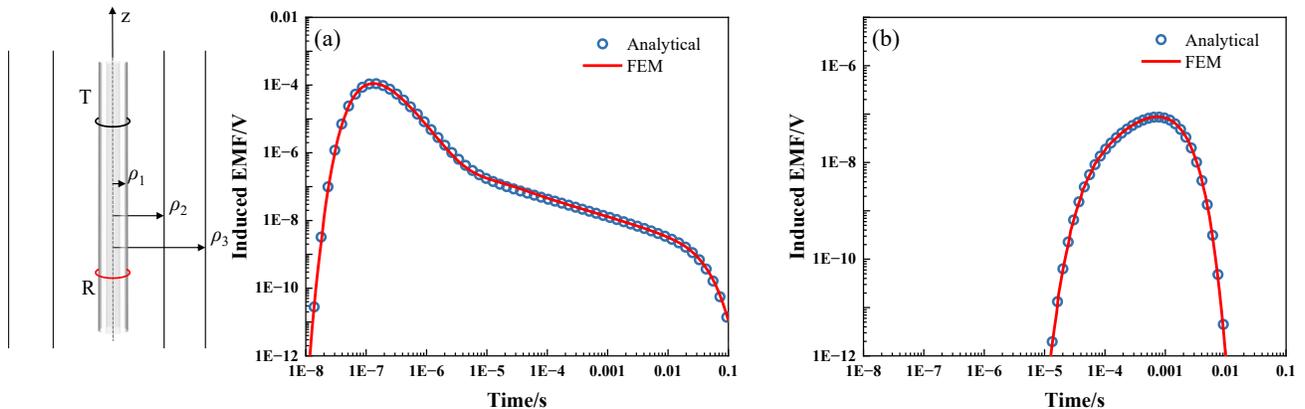

**Fig.5.** Forward modeling in cylindrically multilayered medium

V. CONCLUSION

We present integral methods for simulating transient electromagnetic logging responses through frequency-domain Levin-Sidi extrapolation with time-domain 125-point sine transform for horizontal and cylindrical formation geometries.

For horizontal formations, the solution integrates Hankel transform (using 121-point zeroth-order and 140-point first-order) with the optimized 125-point sine transform, achieving sub-0.1% relative error as demonstrated in homogeneous and layered formations.

The cylindrical formation solution employs Gauss-Legendre spatial quadrature with Levin-Sidi frequency-domain computation, converted via the same 125-point transform, showing remarkable stability in extreme contrast scenarios including cased/open-hole conditions.

Appendix A. Filter coefficients

A1. Sin125: $N = 62$ $s = 0.1490$

| n = -62 to -31 | -30 to 0 | 1 to 31 | 32 to 62 |
|---|---|---|---|
| 2.12901363205731E-11 | 0.00255277086757301 | 0.0781226476856856 | -2.36225355301355E-5 |
| -1.13798756097836E-10 | -0.00423144913832299 | 0.0592912272538349 | 3.60792120642842E-5 |
| 3.57342125808134E-10 | 0.0070790841859956 | 0.0457133447905125 | -2.26720307568542E-5 |
| -8.66177199810141E-10 | -0.0120231715217507 | 0.0339869049092061 | 2.87863704432523E-5 |
| 1.80614817855495E-9 | 0.0209169039150604 | 0.0260834099614931 | -2.07073951006432E-5 |
| -3.43212419998368E-9 | -0.0377375007697752 | 0.0190758044369208 | 2.34384686863551E-5 |
| 6.15312239838286E-9 | 0.0715690924507547 | 0.0147024083924384 | -1.83979987777732E-5 |
| -1.06376386256109E-8 | -0.143628431418173 | 0.0105675450792933 | 1.93591097062032E-5 |
| 1.79824761989623E-8 | 0.300392272973389 | 0.00824880403846011 | -1.60792488145813E-5 |
| -2.99854978142515E-8 | -0.621702887998672 | 0.00579709636307718 | 1.61437686090912E-5 |
| 4.95901051921206E-8 | 1.1596725590709 | 0.00462990476956339 | -1.39050126722959E-5 |
| -8.16119740877693E-8 | -1.68934977435981 | 0.00314977435981177 | 1.35383900845993E-5 |
| 1.33927203494171E-7 | 1.45008135977148 | 0.00261073603801002 | -1.19270888755022E-5 |
| -2.19414399919666E-7 | 0.0783501901661588 | 0.00169082049206614 | 1.13656739956684E-5 |
| 3.59129303794561E-7 | -1.29514846444738 | 0.00148546034023431 | -1.0128487659998E-5 |
| -5.87494730629345E-7 | 0.0967036971475536 | 8.91499107003772E-4 | 9.46813894585851E-6 |
| 9.60785471893266E-7 | 0.933213659926865 | 8.57266824932365E-4 | -8.41664702409356E-6 |
| -1.57099951892435E-6 | 0.505380440501656 | 4.56418116915787E-4 | 7.63915442623139E-6 |
| 2.56853510258259E-6 | -0.21591149914511 | 5.04991841053706E-4 | -6.56674602089751E-6 |
| -4.19926641729636E-6 | -0.55308104991344 | 2.21682460059124E-4 | 5.51953223636671E-6 |
| 6.86516609306759E-6 | -0.508809959840709 | 3.05910460054706E-4 | -4.13462169019483E-6 |
| -1.12234411823352E-5 | -0.285237797314607 | 9.67283891436608E-5 | 2.56451997041369E-6 |
| 1.83486850632148E-5 | -0.0601801185080935 | 1.92056720537216E-4 | -6.78043042556508E-7 |
| -2.99981996685796E-5 | 0.101720756090916 | 3.16515784147247E-5 | -1.1719501448088E-6 |
| 4.90464025529301E-5 | 0.1858096828449 | 1.25807136819039E-4 | 2.52020953880197E-6 |
| -8.01968187352282E-5 | 0.215994458538056 | -9.82775611326966E-7 | -2.56648242635946E-6 |
| 1.31151076101652E-4 | 0.208771484448741 | 8.63191492518657E-5 | 1.0643397698815E-6 |
| -2.1453384276124E-4 | 0.186430612997994 | -1.62180754231696E-5 | 1.10499595252491E-6 |
| 3.51077906725314E-4 | 0.155666256219 | 6.20242591080691E-5 | -2.17168821864917E-6 |
| -5.74936410216533E-4 | 0.127006671869358 | -2.22748516528039E-5 | 1.51770156867007E-6 |
| 9.42657806626588E-4 | 0.0996109111622222 | 4.64799244665232E-5 | -4.08446300711794E-7 |
| -0.00154866178693922 | | | |

REFERENCES

[1] T. J. Cui, W. C. Chew, A. A. Aydiner, D. L. Wright, and D. V. Smith, "Detection of buried targets using a new enhanced very early time electromagnetic (VETEM) prototype system," IEEE Trans. Geosci. Remote Sens., vol. 39, no. 12, pp. 2702–2712, 2001. doi: 10.1109/36.975004

[2] R. S. Smith, A. P. Annan, and P. D. McGowan, "A comparison of data from airborne, semi-airborne, and ground electromagnetic systems," Geophysics, vol. 66, no. 5, pp. 1379-1385, 2001. doi: 10.1190/1.1487084

[3] Z. Xiao et al., "A new fracturing result evaluation method based on borehole time-domain electromagnetic," in Proc. Int. Field Explor. Develop. Conf. 2023, Springer Ser. Geomech. Geoeng., 2024, pp. 454-462, doi: 10.1007/978-981-97-0479-8_40.

[4] T. Hagiwara, "Determination of dip and anisotropy from transient triaxial induction measurements," Geophysics, vol. 77,

no. 4, pp. D105-D112, 2012. doi: 10.1190/geo2011-0503.1

[5] Q. Zhou and D. Gregory, "Investigation on electromagnetic measurement ahead of drill-bit," in Proc. IEEE Int. Geosci. Remote Sens. Symp., 2000, pp. 1745-1747.

[6] F. L. Teixeira and W. C. Chew, "Finite-difference computation of transient electromagnetic waves for cylindrical geometries in complex media," IEEE Trans. Geosci. Remote Sens., vol. 38, no. 4, pp. 1530-1543, 2000.

[7] F. Fu and J. R. Bowler, "Transient eddy current response due to a conductive cylindrical rod," in AIP Conf. Proc., vol. 894, 2007, pp. 332-339. doi:DOI: 10.1063/1.2717991

[8] G. S. Rosa, J. R. Bergmann, and F. L. Teixeira, "Analysis of transient electromagnetic field propagation in well-logging environments via an efficient mode-matching technique," in Proc. Eur. Conf. Antennas Propag., 2019.

[9] T. Theodoros and A. S., "Efficient calculation of transient eddy current response from multilayer cylindrical conductive media," Philos. Trans. R. Soc. A, vol. 378, no. 2182, p. 20190588, Oct. 2020. doi: 10.1098/rsta.2019.0588

[10] J. Li, C. G. Farquharson, and X. Hu, "Three effective inverse Laplace transform algorithms for computing time-domain electromagnetic responses," Geophysics, vol. 81, no. 2, pp. E113-E128, 2016. doi: 10.1190/geo2015-0174.1

[11] F. N. Kong, "Hankel transform filters for dipole antenna radiation in a conductive medium," Geophys. Prospect., vol. 55, no. 1, pp. 83-89, 2007. doi: 10.1111/j.1365-2478.2006.00585.x

[12] F. N. Kong, "Evaluation of Fourier cosine/sine transforms using exponentially positioned samples," J. Appl. Geophys., vol. 79, pp. 46-54, 2012.

[13] K. Key, "Is the fast Hankel transform faster than quadrature?," Geophysics, vol. 77, no. 3, pp. F21-F30, 2012. doi: 10.1190/geo2011-0237.1

[14] D.-C. Hong, S.-W. Wan, H.-M. Liu, N. Li, W. Han, T. Chen, and J.-T. He, "Calculation of tilted coil voltage in cylindrically multilayered medium," in Proc. IEEE Int. Conf. Comput. Electromagn., 2019, pp. 1-3.

[15] Y. Zhao, Z. Zhu, G. Lu, and B. Han, "The optimal digital filters of sine and cosine transforms for geophysical transient electromagnetic method," Journal of Applied Geophysics, vol. 150, pp. 267-277, 2018.

[16] K. A. Michalski and J. R. Mosig, "Efficient computation of Sommerfeld integral tails - methods and algorithms," J. Electromagn. Waves Appl., vol. 30, no. 3, pp. 281-317, 2016.